\documentclass[onecolumn]{article}%
\usepackage{amsmath}
\usepackage{amsfonts}
\usepackage{amssymb}
\usepackage{graphicx}
\usepackage{revsymb}%
\begin{document}

\title{A Systematic Derivation of the Riemannian Barrett-Crane Intertwiner.}
\author{Suresh. K.\ Maran}
\maketitle

\begin{abstract}
The Barrett-Crane intertwiner for the Riemannian$\ $general relativity is
systematically derived by solving the quantum Barrett-Crane constraints
corresponding to a tetrahedron (except for the non-degeneracy condition). It
was shown by Reisenberger that the Barrett-Crane intertwiner is the unique
solution. The systematic derivation can be considered as an alternative proof
of the uniqueness. The new element in the derivation is the rigorous
imposition of the cross-simplicity constraint.

\end{abstract}

\section{Introduction}

A\ quantization of a four-simplex for the Riemannian general relativity was
proposed by Barrett and Crane \cite{BCReimmanion}. It was built on the idea of
the Barrett-Crane intertwiner. It was shown by Reisenberger
\cite{ReisenBCinter} that the Barrett-Crane intertwiner is the unique solution
to the Barrett-Crane constraints corresponding a tetrahedron (except for the
non-degeneracy condition). Here I would I\ like to present an alternative
proof of uniqueness by systematically deriving the Barrett-Crane intertwiner
by imposing the Barrett-Crane constraints.

\section{Review}

The bivectors $B_{i}$ associated with the ten triangles of a four-simplex in a
flat Riemannian space satisfy the following properties called the
Barrett-Crane constraints \cite{BCReimmanion}:

\begin{enumerate}
\item The bivector changes sign if the orientation of the triangle is changed.

\item Each bivector is simple.

\item If two triangles share a common edge, then the sum of the bivectors is
also simple.

\item The sum of the bivectors corresponding to the edges of any tetrahedron
is zero. This sum is calculated taking into account the orientations of the
bivectors with respect to the tetrahedron.

\item The six bivectors of a four-simplex sharing the same vertex are linearly independent.

\item The volume of a tetrahedron calculated from the bivectors is real and non-zero.
\end{enumerate}

The items two and three can be summarized as follows:
\[
B_{i}\wedge B_{j}=0~\forall i,j,
\]
where $A\wedge B=\varepsilon_{IJKL}A^{IJ}B^{KL}$ and the $i,j$ represents the
triangles of a tetrahedron. If $i=j$, it is referred to as the simplicity
constraint. If $i\neq j$ it is referred as the cross-simplicity constraints.

Barrett and Crane have shown \cite{BCReimmanion} that these constraints are
sufficient to restrict a general set of ten bivectors $E_{b}$ so that they
correspond to the triangles of a geometric four-simplex up to translations and
rotations in a four dimensional flat Riemannian space.

A quantum four-simplex for Riemannian general relativity is defined by
quantizing the Barrett-Crane constraints \cite{BCReimmanion}. The bivectors
$B_{i}$ are promoted to the Lie operators $\hat{B}_{i}$ on the representation
space of the relevant group and the Barrett-Crane constraints are imposed at
the quantum level. The last two constraints are inequalities and they are
difficult to impose. For these reasons here after I\ would like to refer to a
state sum model that satisfies only the first four constraints as an
\textit{essential Barrett-Crane model}, While a state sum model that satisfies
all the six constraints as a \textit{rigorous Barrett-Crane model}. The
Barrett-Crane intertwiner corresponds to essential Barrett-Crane model only.
We will do a systematic derivation of the essential Barrett-Crane model here.

\subsection{The Simplicity Constraint}

Our treatment of the simplicity constraints is basically a review of work done
before \cite{BCReimmanion}, \cite{BFfoamHighD}. The group $SO(4,R)$ is
isomorphic to $\frac{SU(2)\times SU(2)}{Z_{2}}$. An element $B$ of the Lie
algebra of $SO(4)$ can be split into the left and the right handed $SU(2)$
components,%
\begin{equation}
B=B_{L}+B_{R}.
\end{equation}
There are two Casimir operators for $SO(4)$ which are
\begin{align*}
&  \varepsilon_{IJKL}B^{IJ}B^{KL}~~\text{and}\\
&  \delta_{IK}\delta_{JL}B^{IJ}B^{KL},
\end{align*}
where $\eta_{IK}$ is the flat Euclidean metric. In terms of the left and right
handed split I can expand the Casimir operators as%
\[
\varepsilon_{IJKL}B^{IJ}B^{KL}=B_{L}\cdot B_{L}-B_{R}\cdot B_{R}~~\text{and}%
\]%
\[
\delta_{IK}\delta_{JL}B^{IJ}B^{KL}=B_{L}\cdot B_{L}+B_{R}\cdot B_{R},
\]
where the dot products are the trace in the $SU(2)$ Lie algebra coordinates.

The bivectors are to be quantized by promoting the Lie algebra vectors to Lie
operators on the unitary representation space of $SO(4)\cong$ $\frac
{SU(2)\times SU(2)}{Z_{2}}$. The relevant unitary representations of $SO(4)$
are labeled by a pair ($J_{L},$ $J_{R}$) of unitary representations of
$SU(2)$. The elements of the representation space $D_{J_{L}}\otimes$
$D_{J_{R}}$ are the eigen states of the Casimirs and on them the operators
reduce to the following:
\begin{equation}
\varepsilon_{IJKL}\hat{B}^{IJ}\hat{B}^{KL}=\frac{J_{L}(J_{L}+1)-J_{R}%
(J_{R}+1)}{2}\hat{I}~~\text{and} \label{eq.1}%
\end{equation}%
\begin{equation}
\delta_{IK}\delta_{JL}\hat{B}^{IJ}\hat{B}^{KL}=\frac{J_{L}(J_{L}%
+1)+J_{R}(J_{R}+1)}{2}\hat{I}. \label{eq.2}%
\end{equation}
The equation (\ref{eq.1}) implies that on $D_{J_{L}}\otimes$ $D_{J_{R}}$ the
simplicity constraint $B\wedge B=0$ is equivalent to the condition
$J_{L}=J_{R}$. I would like to find a representation space on which the
representations of $SO(4)$ are restricted precisely by $J_{L}$ $=$ $J_{R}$.

In Ref:\cite{BFfoamHighD} it has been shown for $SO(N,R)$ that the simplicity
constraint reduces the Hilbert space associated to a triangle to that of the
$L^{2}$ functions on $S^{N-1}$. Consider a square integrable function $f$
$(x)$ on the sphere $S^{3}$ defined by%

\[
x\cdot x=1,\forall x\in\boldsymbol{R}^{4}.
\]
It can be Fourier expanded in the representation matrices of $SU(2)$ using the
isomorphism $S^{3}\cong$ $SU(2)$,%
\begin{equation}
f(x)=\sum\limits_{J}d_{J}Tr(F_{J}T_{J}(\mathfrak{g}(x)^{-1}), \label{Cs3Exp}%
\end{equation}
where $\mathfrak{g}\mathrm{:}S^{3}$ $\longrightarrow SU(2)$ is an isomorphism
from $S^{3}$ to $SU(2),$ $F_{m_{2}J}^{m_{1}}$ the Fourier coefficients,
$T_{m_{2}J}^{m_{1}}(g)$ are the matrix elements of spin $J$ representation of
an element $g\in SU(2)$ and $d_{J}$ the dimension of the $J$ representation.
The group action of $g=(g_{L},g_{R})\in SO(4)$ on $x$ $\in S^{3}$ is given by
\begin{equation}
\mathfrak{g}(gx)=g_{L}^{-1}\mathfrak{g}(x)g_{R}. \label{Cs3Action}%
\end{equation}
Using equation (\ref{Cs3Exp}) I can consider the $T_{J}(\mathfrak{g}%
(x))(m_{1},m_{2})$ as the basis functions of $L^{2}$ functions on $S^{3}.$ The
matrix elements of the action of $g$ on $S^{3}$ is given by%
\[
\int\bar{T}_{\acute{m}_{2}\acute{J}}^{\acute{m}_{1}}(\mathfrak{g}%
(x))T_{m_{2}J}^{m_{1}}(\mathfrak{g}(gx))dx=\frac{1}{d_{J}}\bar{T}_{m_{1}%
\acute{J}}^{\acute{m}_{1}}(g_{L})T_{m_{2}J}^{\acute{m}_{2}}(g_{R}%
)\delta(\acute{J}-J).
\]
I see that the representation matrices are precisely those of $SO(4)$ only
restricted by the constraint $J_{L}=J_{R}$. Since we have all the simple
representations included in the representation, the simplicity constraint
effectively reduces the Hilbert space $H$ to the space of $L^{2}$ functions on
$S^{3}$.

\subsection{The Cross-Simplicity Constraints}

Next let me quantize the cross-simplicity constraint part of the Barrett-Crane
constraint. Consider the quantum state space associated with a pair of
triangles $1$ and $2$ of a tetrahedron. A general quantum state that just
satisfies the simplicity constraints $B_{1}\wedge B_{1}=0$ and $B_{2}\wedge
B_{2}=0$ is of the form $f(x_{1},x_{2})$ $\in L^{2}(S^{3}\times S^{3})$,
$x_{1},x_{2}\in S^{3}$.

On the elements of $L^{2}(S^{3}\times S^{3})$ the action $B_{1}\wedge B_{2}$
is equivalent to the action of $\left(  B_{1}+B_{2}\right)  \wedge\left(
B_{1}+B_{2}\right)  $\footnote{Please notice that
\[
\left(  \hat{B}_{1}+\hat{B}_{2}\right)  \wedge\left(  \hat{B}_{1}+\hat{B}%
_{2}\right)  =\hat{B}_{1}\wedge\hat{B}_{1}+\hat{B}_{2}\wedge\hat{B}_{2}%
+2B_{1}\wedge\hat{B}_{2}.
\]
But since $\hat{B}_{1}\wedge\hat{B}_{1}=\hat{B}_{2}\wedge\hat{B}_{2}=0$ on
$f(x_{1},x_{2})$ we have
\[
\left(  \hat{B}_{1}+\hat{B}_{2}\right)  \wedge\left(  \hat{B}_{1}+\hat{B}%
_{2}\right)  f(x_{1},x_{2})=\hat{B}_{1}\wedge\hat{B}_{2}f(x_{1},x_{2}).
\]
}. This implies that the cross-simplicity constraint $B_{1}\wedge B_{2}=0$
requires the simultaneous rotation of $x_{1},x_{2}$ involve only the $J_{L}$
$=J_{R}$ representations. The simultaneous action of $g=(g_{L},g_{R})$ on the
arguments of $f(x_{1},x_{2})$ is%
\begin{equation}
gf(x_{1},x_{2})=f(g_{L}^{-1}x_{1}g_{R},g_{L}^{-1}x_{2}g_{R}). \label{action}%
\end{equation}
The harmonic expansion of $f(x_{1},x_{2})$ in terms of the basis function
$T_{J}(\mathfrak{g}(x))(m_{1},m_{2})$ is%
\[
f(x_{1},x_{2})=\sum\limits_{J}F_{m_{1}m_{2}J_{1}J_{2}}^{\acute{m}_{1}\acute
{m}_{2}}T_{\acute{m}_{1}J_{1}}^{m_{1}}(\mathfrak{g}(x_{1}))T_{\acute{m}%
_{2}J_{2}}^{m_{2}}(\mathfrak{g}(x_{2})).
\]
The rest of the calculations can be understood graphically. The last equation
can be graphically written as follows:%

\[
f(x_{1},x_{2})=\sum\limits_{J_{1}J_{2}}%
\raisebox{-0.4021in}{\includegraphics[
height=0.8518in,
width=1.2341in
]%
{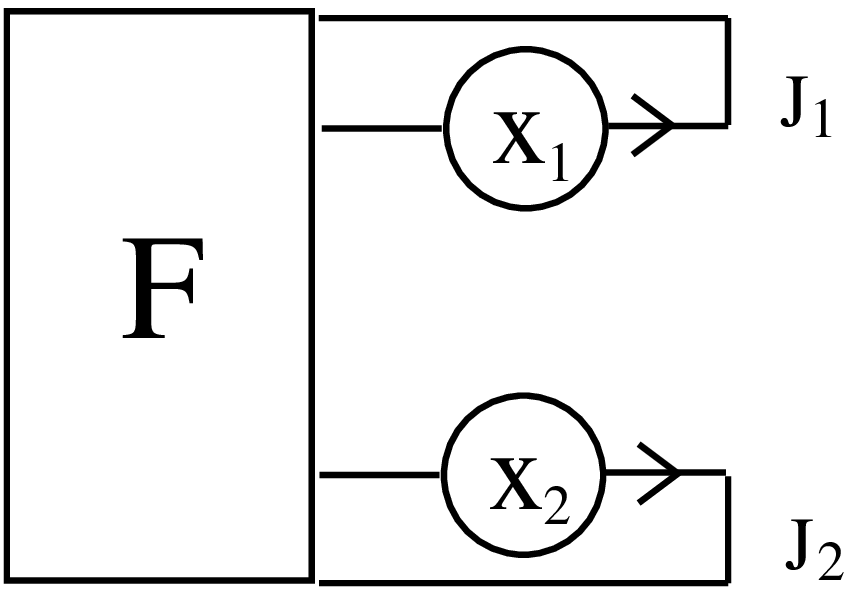}%
}%
,
\]
where the box $F$ represents the Fourier coefficient $F_{m_{1}m_{2}J_{1}J_{2}%
}^{\acute{m}_{1}\acute{m}_{2}}$. The action of $g\in SO(4)$ on $f$ is%

\begin{equation}
gf(x_{1},x_{2})=\sum\limits_{J_{1}J_{2}}%
\raisebox{-0.4021in}{\includegraphics[
height=0.832in,
width=1.9061in
]%
{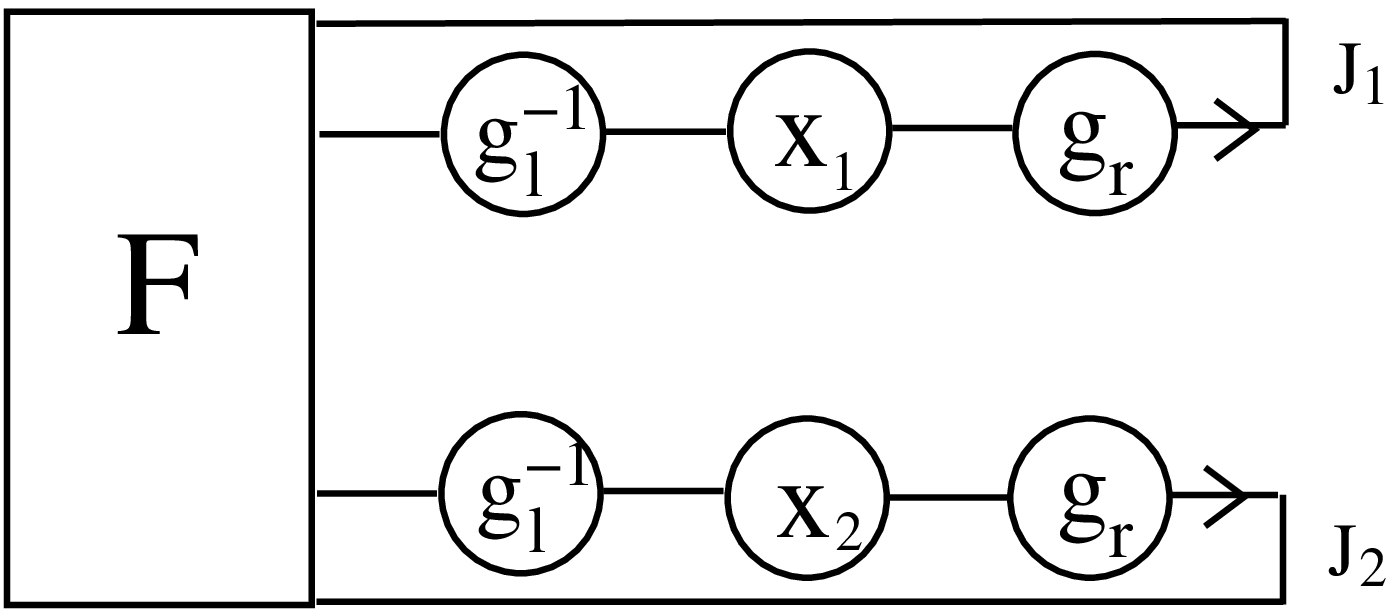}%
}%
. \label{gfresult}%
\end{equation}
Now for any $h\in SU(2),$%
\[
T_{a_{1}J_{1}}^{b_{1}}(h)T_{a_{2}J_{1}}^{b_{2}}(h)=\sum_{J_{3}}C_{J_{1}%
J_{2}b_{3}}^{b_{1}b_{2}J_{3}}\bar{C}_{a_{1}a_{2}J_{3}}^{J_{1}J_{2}a_{3}%
}T_{a_{3}J_{3}}^{b_{3}}(h),
\]
where $C$'s are the Clebsch-Gordan coefficients of $SU(2)$ \cite{vmk}. I have
assumed all the repeated indices are either integrated or summed over for the
previous and the next two equations. Using this I can rewrite the $g_{L}$ and
$g_{R}$ parts of the result (\ref{gfresult}) as follows:%
\begin{equation}
T_{a_{1}J_{1}}^{m_{1}}(g_{L}^{-1})T_{a_{2}J_{2}}^{m_{2}}(g_{L}^{-1}%
)=\sum_{J_{L}}C_{J_{1}J_{2}m_{3}}^{m_{1}m_{2}J_{L}}\bar{C}_{a_{1}a_{2}J_{L}%
}^{J_{1}J_{2}a_{3}}T_{a_{3}J_{L}}^{m_{3}}(g_{L}^{-1}) \label{exp1}%
\end{equation}
and%
\begin{equation}
T_{\acute{m}_{1}J_{1}}^{b_{1}}(g_{R})T_{\acute{m}_{2}J_{2}}^{b_{2}}%
(g_{R})=\sum_{J_{L}}C_{J_{1}J_{2}b_{3}}^{b_{1}b_{2}J_{R}}\bar{C}_{\acute
{m}_{1}\acute{m}_{2}J_{R}}^{J_{1}J_{2}\acute{m}_{3}}T_{\acute{m}_{3}J_{R}%
}^{b_{3}}(g_{R}). \label{exp2}%
\end{equation}
Now we have
\[
gf(x_{1},x_{2})=\sum\limits_{J_{1}J_{2}J_{L}J_{R}}%
\raisebox{-0.4516in}{\includegraphics[
height=0.9381in,
width=1.9535in
]%
{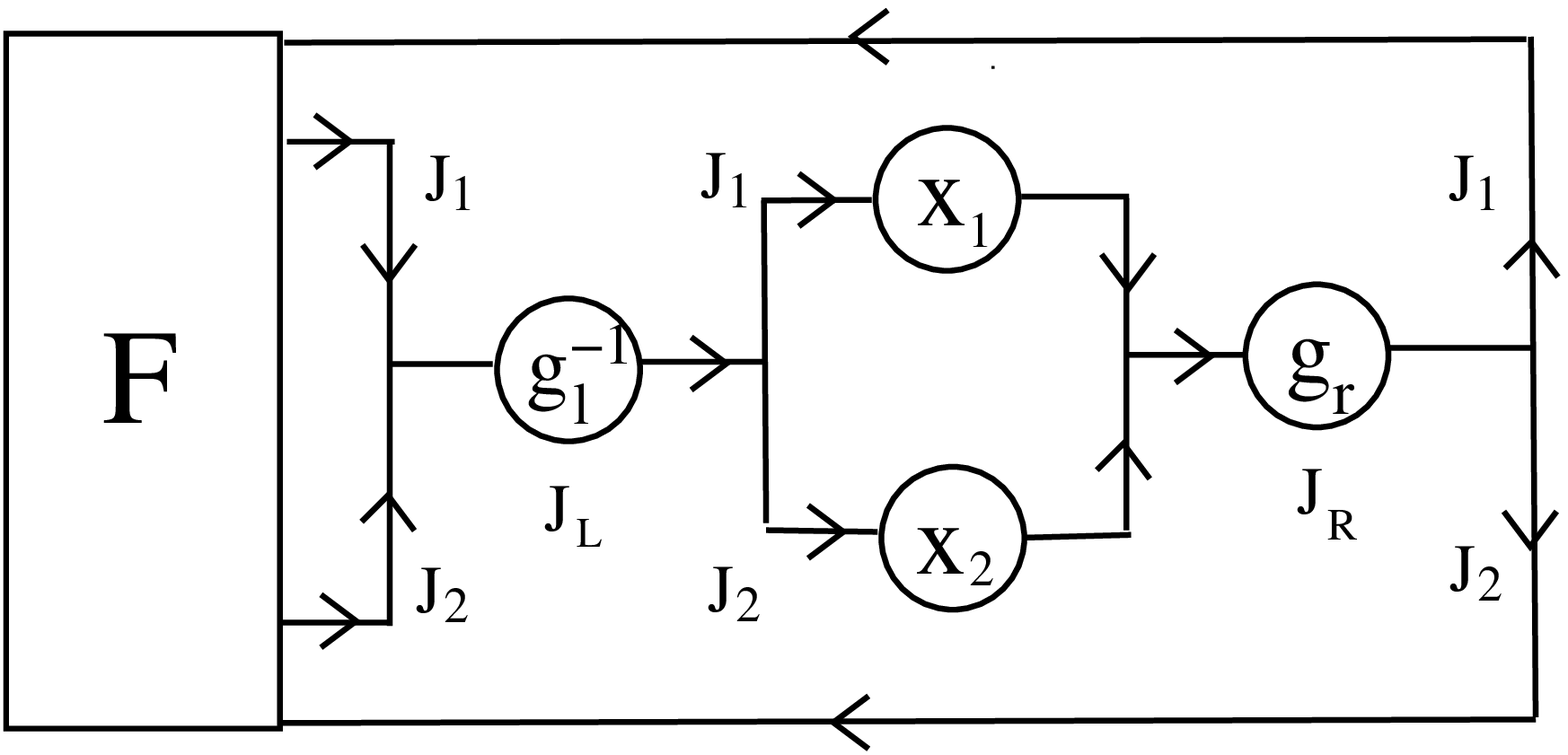}%
}%
.
\]

To satisfy the cross-simplicity constraint the expansion of $gf(x_{1},x_{2})$
must have contribution only from the terms with $J_{L}=J_{R}$. Let me remove
all the terms which do not satisfy $J_{L}=J_{R}$. Also let me set $g=I.$ Now
we can deduce that the functions denoted by $\tilde{f}(x_{1},x_{2})$ obtained
by reducing $f(x_{1},x_{2})$ using the cross-simplicity constraints must have
the expansion,%

\begin{equation}
\tilde{f}(x_{1},x_{2})=\sum\limits_{J_{1}J_{2}J}c_{J}%
\raisebox{-0.5016in}{\includegraphics[
height=0.9729in,
width=2.0928in
]%
{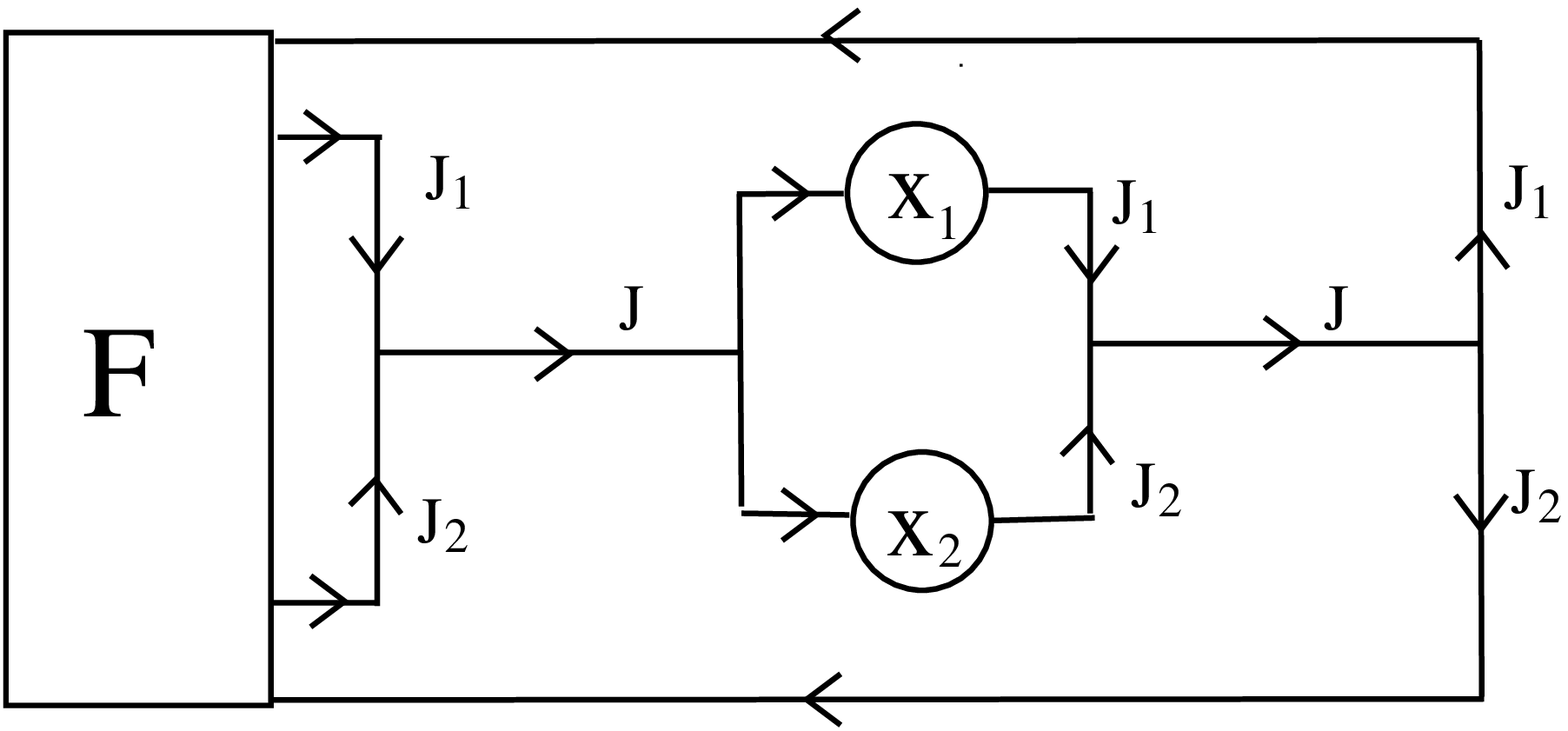}%
}%
\text{,}%
\end{equation}
where the coefficients $c_{J}$ introduced depends on the precise definition of
the cross-simplicity projector. But as we will see, the final answer does not
depend on the $c_{J}$'s. Now the Clebsch-Gordan coefficient terms in the
expansion can be re-expressed using the following equation :%
\begin{equation}
C_{J_{1}J_{2}m_{3}}^{m_{1}m_{2}J}\bar{C}_{\acute{m}_{1}\acute{m}_{2}J}%
^{J_{1}J_{2}\acute{m}_{3}}=\frac{1}{d_{J}}\int_{SU(2)}T_{\acute{m}_{1}J_{1}%
}^{m_{1}}(h)T_{\acute{m}_{2}J_{2}}^{m_{2}}(h)\bar{T}_{m_{3}J}^{\acute{m}_{3}%
}(h)dh,\label{clebsmpl}%
\end{equation}
where $h$, $\tilde{h}$ $\in$ $SU(2)$ and $dh$ the bi-invariant measure on
$SU(2)$. Using this in the two middle Clebsch-Gordan coefficients of
$\tilde{f}(x_{1},x_{2})$ we get
\[
\tilde{f}(x_{1},x_{2})=\sum\limits_{J_{1}J_{2}J}\int_{SU(2)}\frac{c_{J}}%
{d_{J}}%
\raisebox{-0.5571in}{\includegraphics[
height=1.2013in,
width=1.9419in
]%
{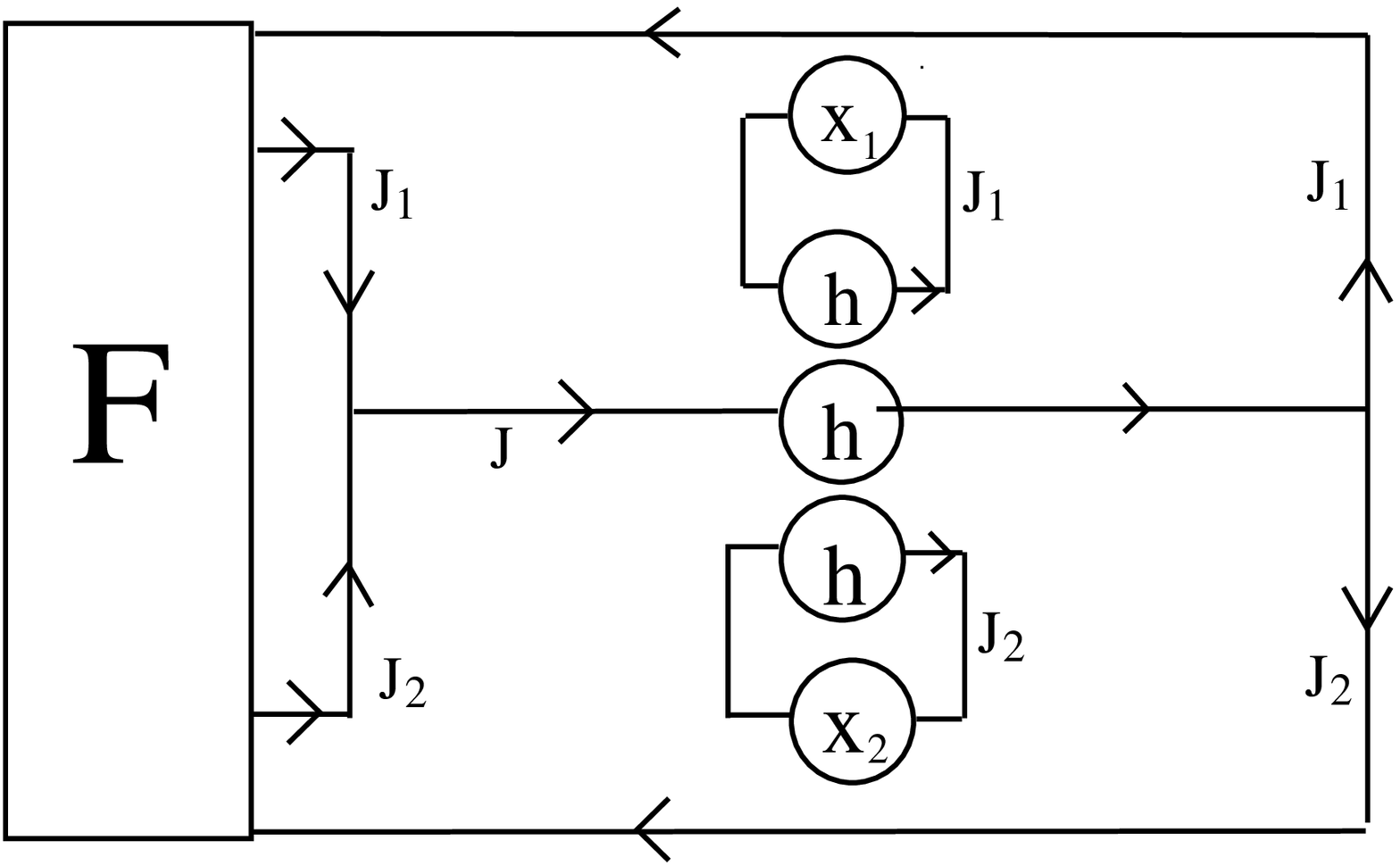}%
}%
dh.
\]
This result can be rewritten for clarity as%
\[
\tilde{f}(x_{1},x_{2})=\sum\limits_{J_{1}J_{2}J}\int_{SU(2)}\frac{c_{J}}%
{d_{J}}%
\raisebox{-0.6102in}{\includegraphics[
height=1.2312in,
width=1.6496in
]%
{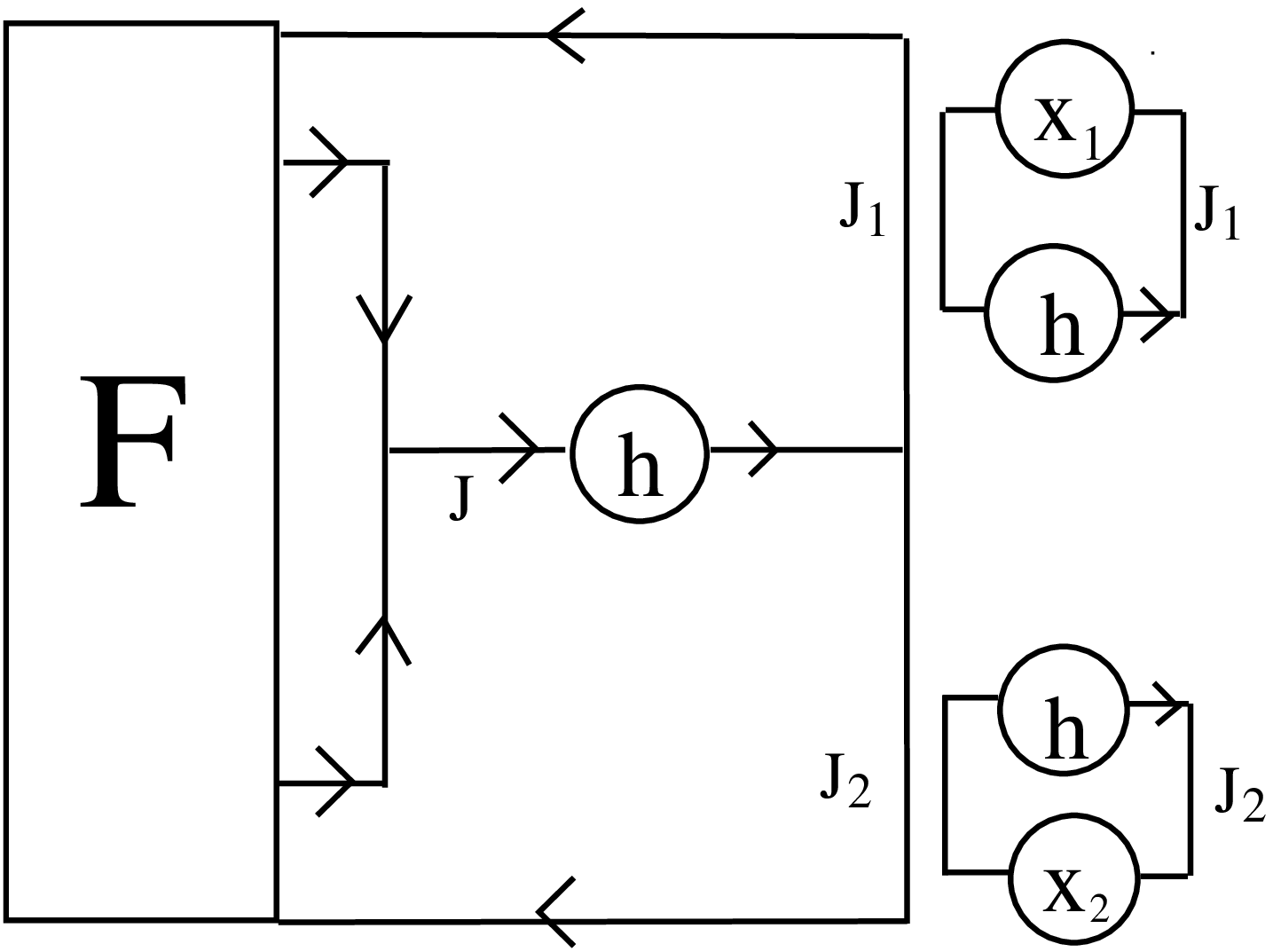}%
}%
dh.
\]
Once again applying equation (\ref{clebsmpl}) to the remaining two
Clebsch-Gordan coefficients we get,%

\[
\tilde{f}(x_{1},x_{2})=\sum\limits_{J_{1}J_{2}J}c_{J}\iint_{SU(2)\times
SU(2)}\left(  \frac{1}{d_{J}}\right)  ^{2}%
\raisebox{-0.5562in}{\includegraphics[
height=1.2254in,
width=1.5691in
]%
{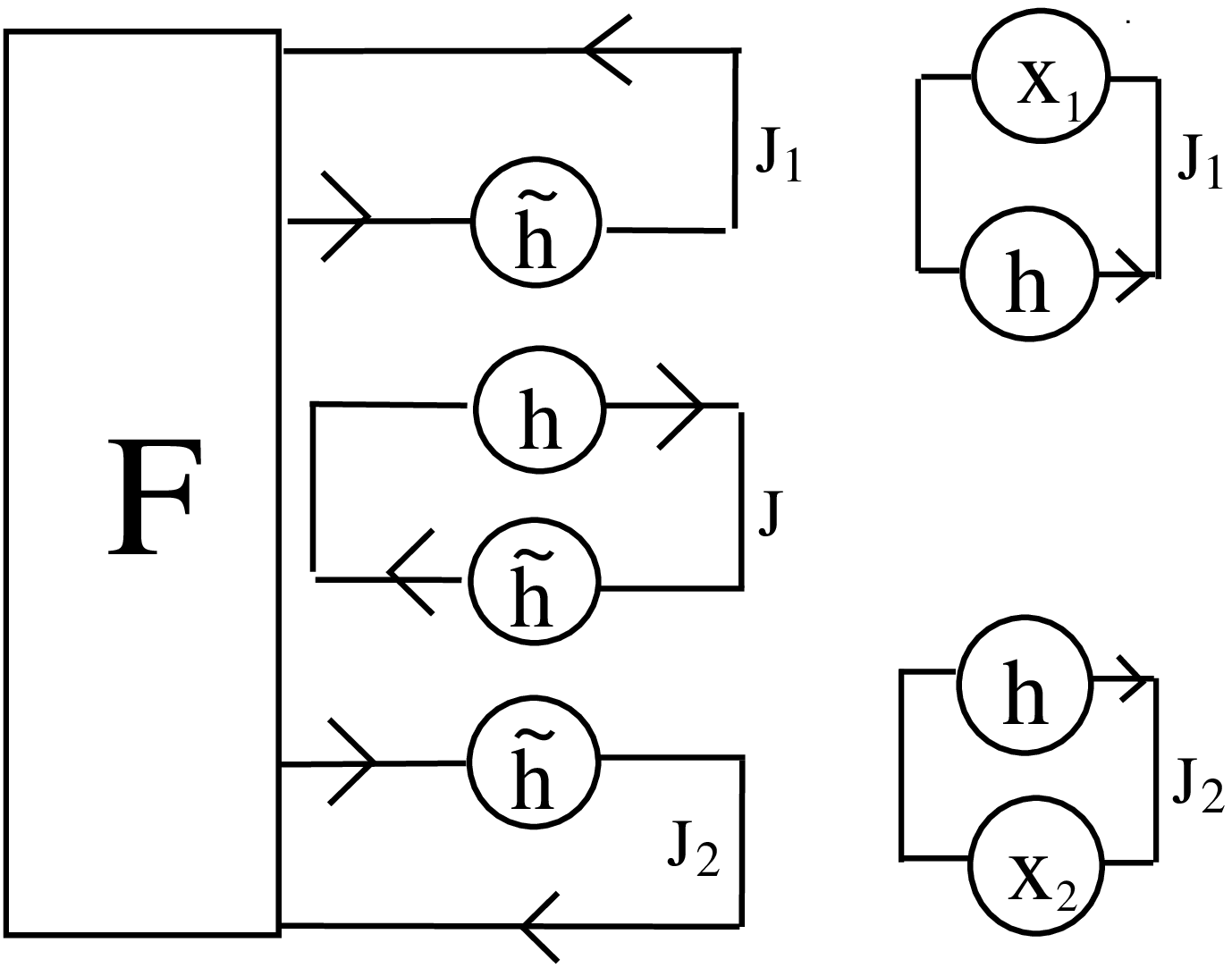}%
}%
~d\tilde{h}dh.
\]

By rewriting the above expression, I deduce that a general function $\tilde
{f}(x_{1},x_{2})$ that satisfies the cross-simplicity constraint must be of
the form,
\begin{subequations}
\label{xsimpfunc}%
\begin{align*}
\tilde{f}(x_{1},x_{2})  &  =\sum\limits_{J_{1}J_{2}}\int_{SU(2)}F_{J_{1}J_{2}%
}(h)%
\raisebox{-0.3009in}{\includegraphics[
height=0.5863in,
width=1.337in
]%
{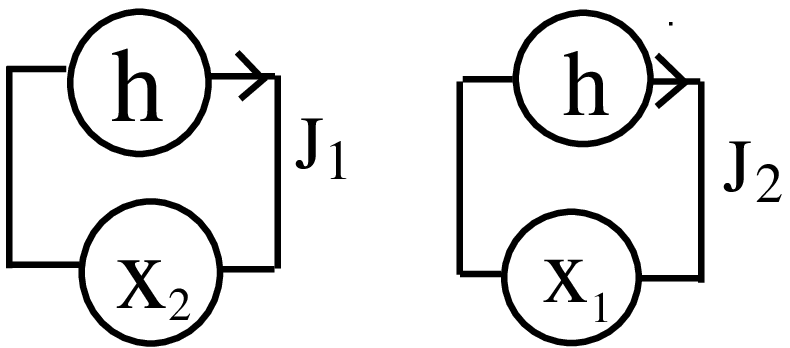}%
}%
dh,\\
&  =\sum\limits_{J_{1}J_{2}}\int_{SU(2)}F_{J_{1}J_{2}}(h)tr(T_{J_{1}%
}(\mathfrak{g}(x_{1})h)tr(T_{J_{2}}(\mathfrak{g}(x_{2})h)dh,
\end{align*}
where $F_{J_{1}J_{2}}(h)$ is arbitrary.

If $\Psi(x_{1},x_{2},x_{3},x_{4})$ is the quantum state of a tetrahedron that
satisfies all of the simplicity constraints and the cross-simplicity
constraints, it must be of the form,
\end{subequations}
\begin{align*}
&  \Psi(x_{1},x_{2},x_{3},x_{4})\\
&  =\sum_{J_{1}J_{2}J_{3}J_{4}}F_{J_{1}J_{2}J_{3}J_{4}}(h)tr(T_{J_{1}%
}(\mathfrak{g}(x_{1})h)tr(T_{J_{2}}(\mathfrak{g}(x_{2})h)\\
&  tr(T_{J_{3}}(\mathfrak{g}(x_{3})h)tr(T_{J_{4}}(\mathfrak{g}(x_{4})h)dh.
\end{align*}
This general form is deduced by requiring that for every pair of variables
with the other two fixed, the function must be the form of the right hand side
of equation (\ref{xsimpfunc}).

\subsection{The $SO(4)$ Barrett-Crane Intertwiner}

Now the quantization of the fourth Barrett-Crane constraint demands that
$\Psi$ is invariant under the simultaneous rotation of its variables. This is
achieved if $F_{J_{1}J_{2}J_{3}J_{4}}(h)$ is a constant function of $h$.
Therefore the quantum state of a tetrahedron is spanned by%
\begin{equation}
\Psi_{J_{1}J_{2}J_{3}J_{4}}(x_{1},x_{2},x_{3},x_{4})=\int_{n\in S^{3}}%
{\displaystyle\prod\limits_{i}}
T_{J_{i}}(\mathfrak{g}(x_{i})\mathfrak{g}(n))dn, \label{BCintertwiner}%
\end{equation}
where the measure $dn$ on $S^{3}$ is derived from the bi-invariant measure on
$SU(2)$.

The quantum state can be diagrammatically represented as follows:%
\[
\Psi_{J_{1}J_{2}J_{3}J_{4}}(x_{1},x_{2},x_{3},x_{4})=\int%
\raisebox{-0.66in}{\includegraphics[
height=1.3375in,
width=1.3375in
]%
{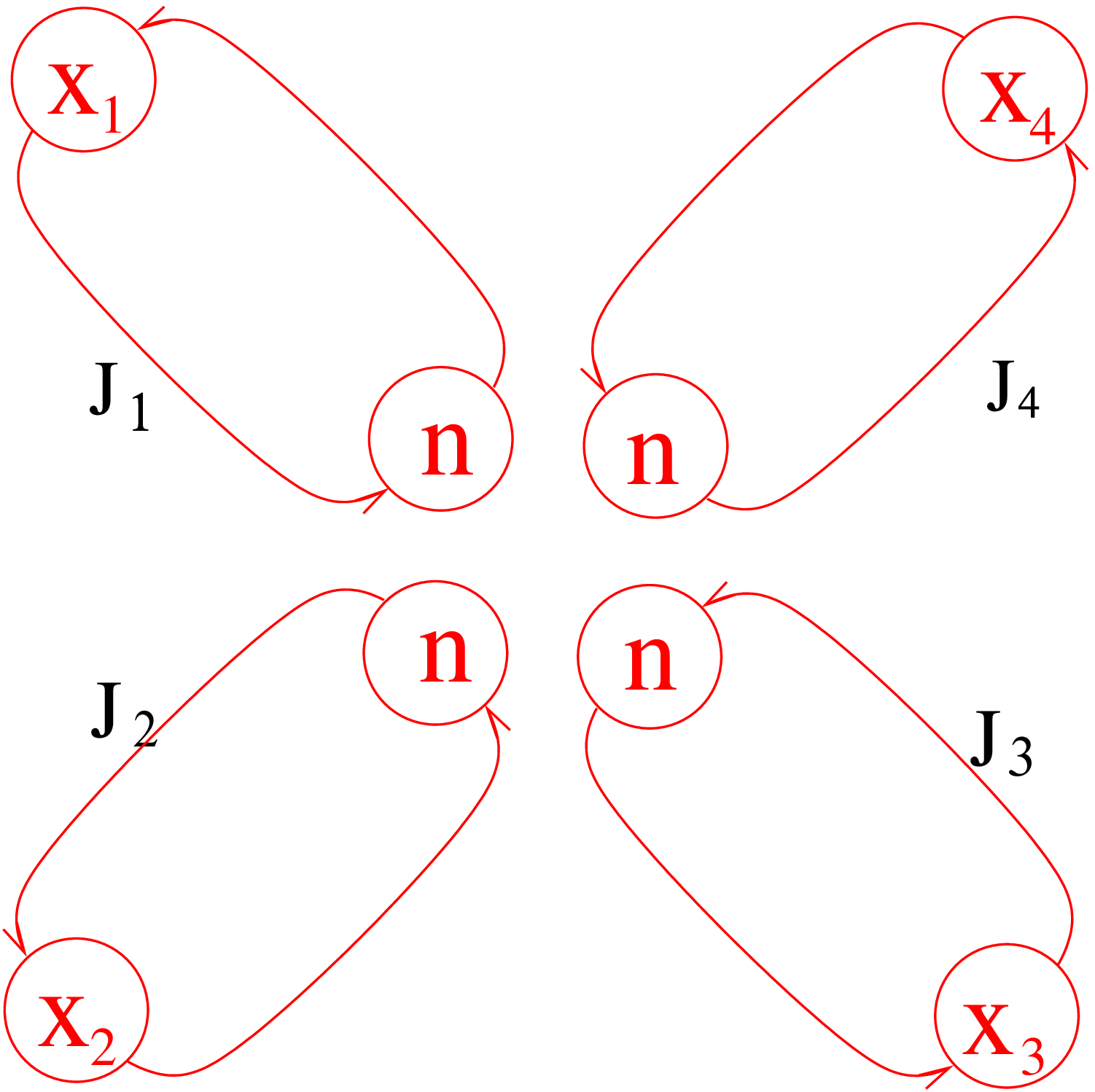}%
}%
dn.
\]
A unitary representation $T_{J}$ of $SU(2)$ can be considered as an element of
$D_{J}\otimes D_{J}^{\ast}$ where $D_{J}^{\ast}$ is the dual representation of
$D_{J}$. So using this the Barrett-Crane intertwiner can be written as an
element $\left\vert \Psi_{J_{1}J_{2}J_{3}J_{4}}\right\rangle \in
\bigotimes\limits_{i}D_{J_{i}}\otimes D_{J_{i}}^{\ast}$ as follows:%
\[
\left\vert \Psi_{J_{1}J_{2}J_{3}J_{4}}\right\rangle =\int\limits_{S^{3}}%
\raisebox{-0.6584in}{\includegraphics[
height=1.2777in,
width=1.2943in
]%
{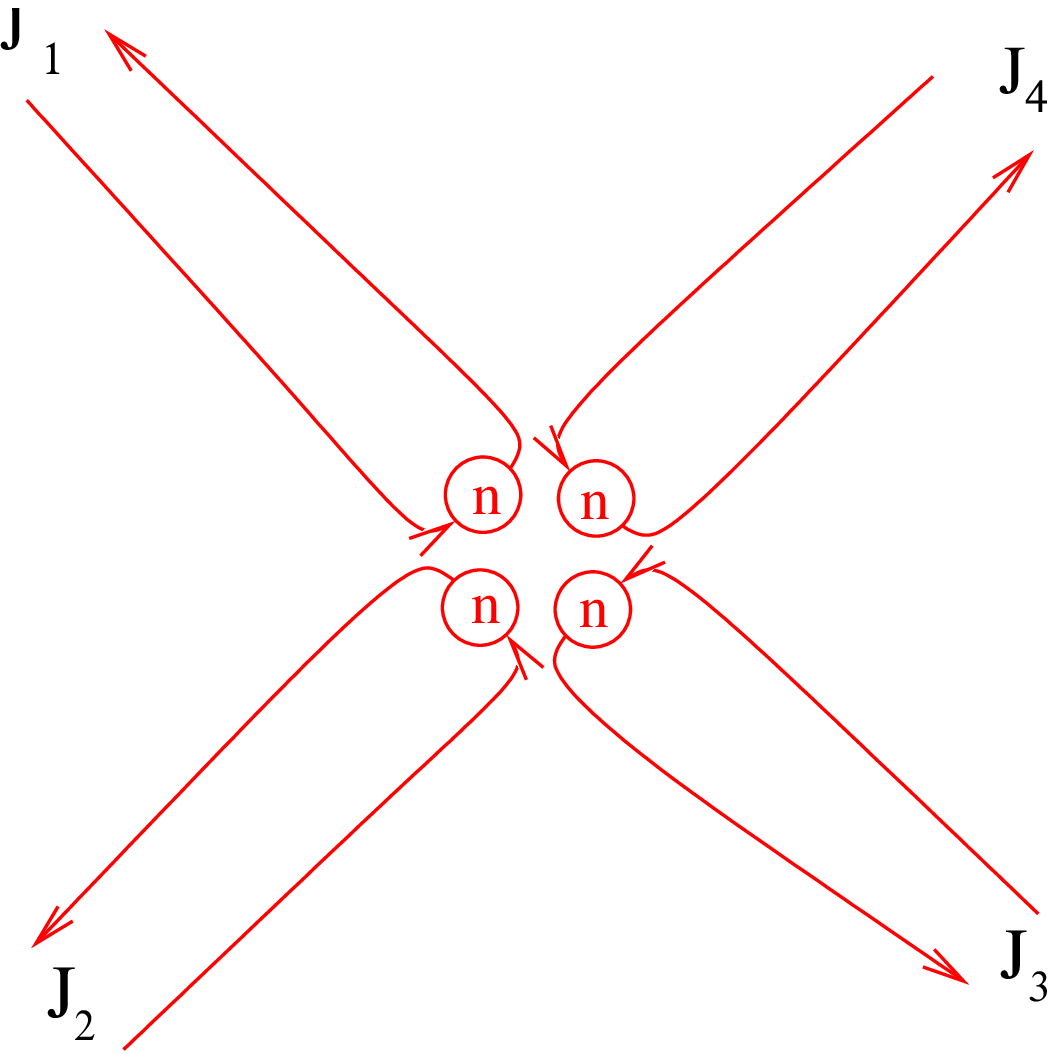}%
}%
dn.
\]
Since $SU(2)\approx S^{3},$ using the following graphical identity:%

\[
\int_{SU(2)}%
\raisebox{-0.5518in}{\includegraphics[
height=1.1191in,
width=1.0594in
]%
{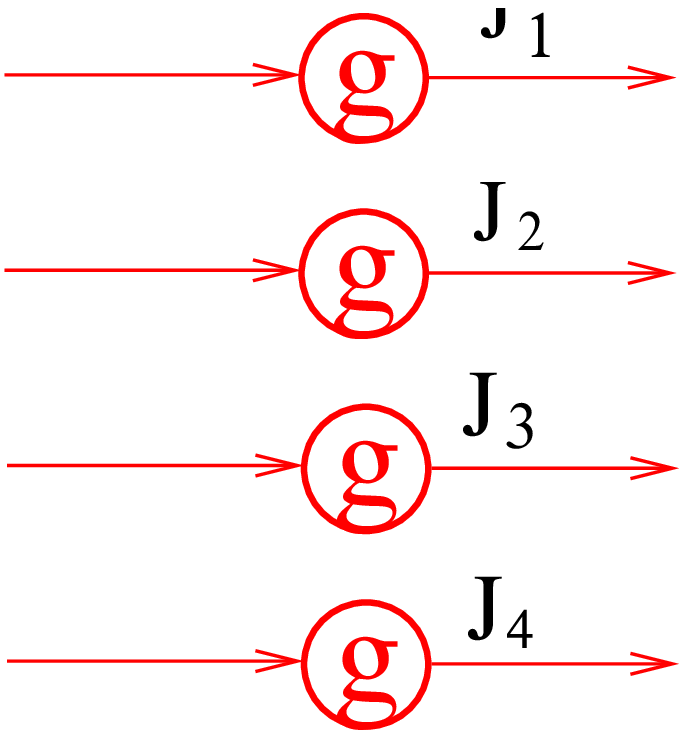}%
}%
dg=\sum_{J}\frac{1}{d_{J}}%
\raisebox{-0.5016in}{\includegraphics[
height=1.0334in,
width=1.8447in
]%
{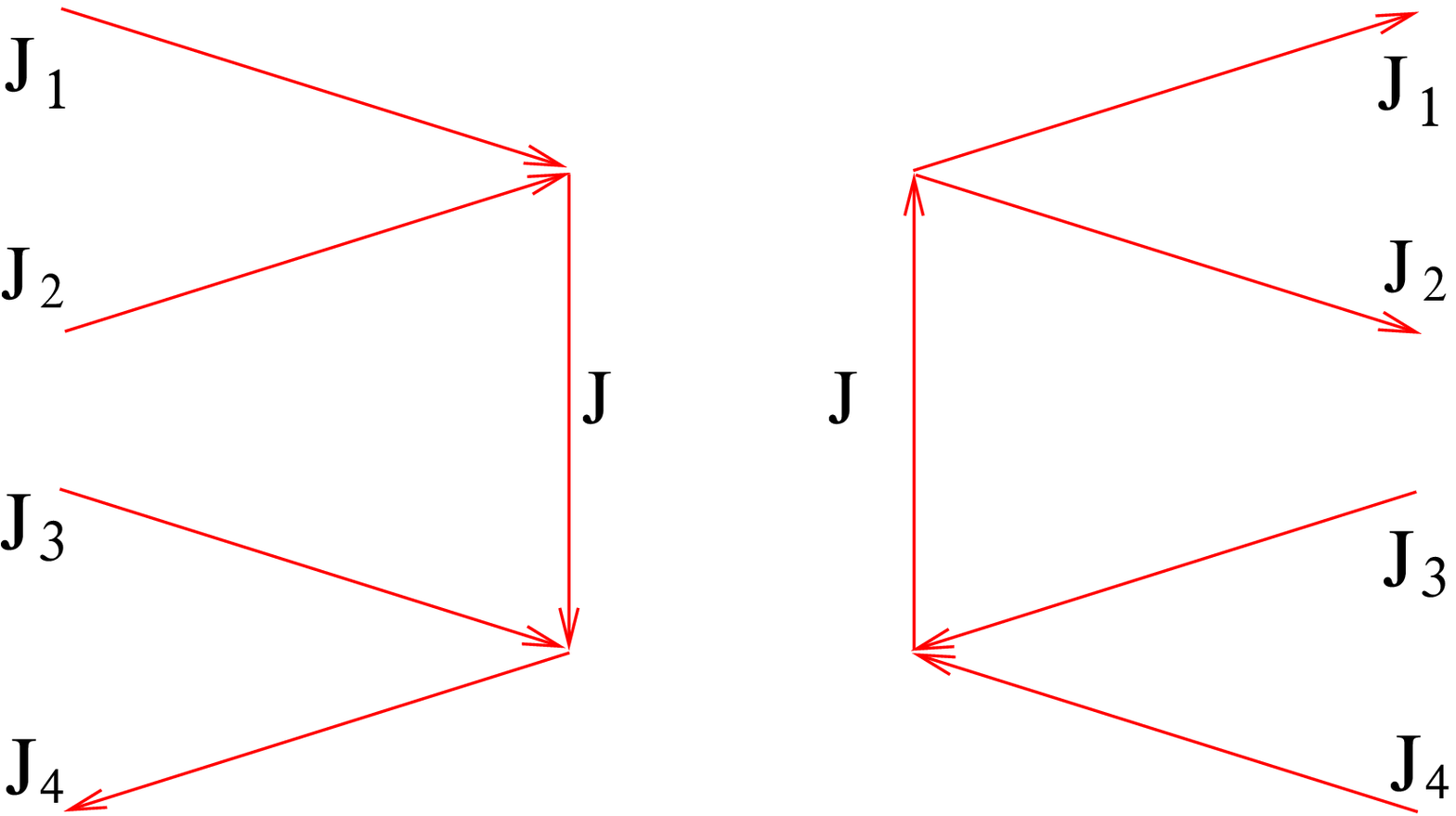}%
}%
,
\]
the Barrett-Crane solution can be rewritten as%

\[
\left\vert \Psi_{J_{1}J_{2}J_{3}J_{4}}\right\rangle =\sum_{J}\frac{1}{d_{J}}%
\raisebox{-0.4541in}{\includegraphics[
height=0.9921in,
width=1.8074in
]%
{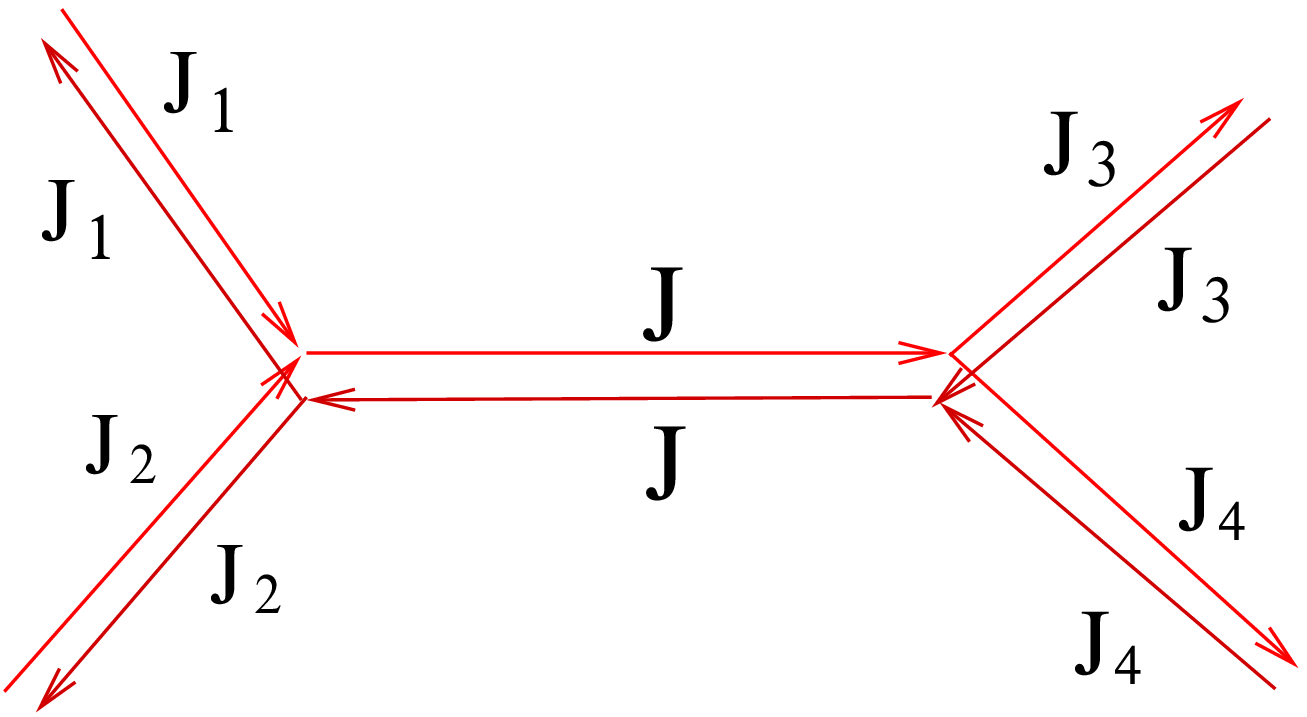}%
}%
,
\]
which emerges as an intertwiner in the familiar form in which Barrett and
Crane proposed it for the Riemannian general relativity.


\begin{thebibliography}{9}                                                                                                %


\bibitem {BCReimmanion}J. W. Barrett and L. Crane, Relativistic Spin Networks
and Quantum general relativity. J.Math.Phys., 39:3296--3302, 1998.

\bibitem {ReisenBCinter}M. P. Reisenberger, On Relativistic Spin Network
Vertices, J.Math.Phys., 40:2046--2054, 1999.

\bibitem {BFfoamHighD}L. Freidel, K. Krasnov and R. Puzio, BF Description of
Higher-Dimensional Gravity Theories, Adv. Theor. Math. Phys. 3 (1999)
1289-1324, arXiv:hep-th/9901069.

\bibitem {vmk}D. A. Varshalovich, A. N. Moskalev and V. K. Khersonskii,
Quantum Theory of Angular Momentum, World Scientific, 1988.
\end{thebibliography}
\end{document}